\documentclass[preprint,12pt]{elsarticle}
\usepackage[utf8]{inputenc}
\usepackage[english]{babel}
 
\usepackage[dvipsnames]{xcolor}

\usepackage{mathenv}

\usepackage{color}

\definecolor{myhigh}{HTML}{228b22}
\definecolor{myhighB}{HTML}{6a5acd}

\usepackage{amssymb}

\journal{Physics Letters B}

\begin{document}
\def\spa#1.#2{\left\langle#1\,#2\right\rangle}
\def\spb#1.#2{\left[#1\,#2\right]}

\def\tree{\rm tree}

\def\la{\langle}
\def\ra{\rangle}

\def\Mloop{M^{\oneloop}}
\def\Mtree{M^{\tree}}
\def\Atree{A^{\tree}}
\def\Aloop{A^{\oneloop}}

\begin{frontmatter}



\title{Diagrammar in an Extended Theory of Gravity}


\author{David~C.~Dunbar, John~H.~Godwin, Guy~R.~Jehu and Warren~B.~Perkins}

\address{College of Science, \\
Swansea University, \\
Swansea, SA2 8PP, UK}

\begin{abstract}
We show how the $S$-matrix of an extended theory of gravity defined by its three-point amplitudes can be constructed by demanding factorisation.  The 
resultant $S$-matrix has tree amplitudes obeying the same soft singularity theorems as Einstein gravity including the sub-sub-leading terms. 
\end{abstract}


\end{frontmatter}


\section{Introduction}

Scattering amplitudes are traditionally defined from a quantum field theory and the resulting  Feynman vertices and Feynman diagrams. 
Alternatively, the amplitudes can be regarded as the fundamental objects which define the theory perturbatively.
It is not very useful to define a theory by specifying
the entire $S$-matrix explicitly but it is an important question whether the $S$-matrix can be defined from a minimal set of data and rules
i.e. a {\it "diagrammar"}~\cite{tHooft:1973wag}.   
Once a minimal set of amplitudes is specified we aim to construct all other amplitudes by demanding they have the correct symmetries
and singularities.  
Defining the $S$-matrix using its singularities is a long-standing programme which is still 
active and fruitful~\cite{Eden,Bern:1994zx,Bern:1994cg,Britto:2005fq,Arkani-Hamed:2013jha,Arkani-Hamed:2016rak}.

In this letter we build an $S$-matrix from a set of three-point amplitudes using their singularity structure.  
The $S$-matrix corresponds to a theory of  Einstein gravity extended by the addition of $R^3$ terms.  
We are working with massless theories and view the amplitude as a function of the twistor variables $\lambda_i^{a}$ and $\bar\lambda_i^{\dot a}$, $M(\lambda_i,\bar\lambda_i)$.  The spinor products $\spa{i}.j,\spb{i}.j$ are 
$\spa{i}.j=\epsilon_{ab} \lambda^a_i\lambda^b_j$, $\spb{i}.j=\epsilon_{\dot a\dot b} \bar\lambda^{\dot a}_i\bar\lambda^{\dot b}_j$.
In this formalism amplitudes have a well-defined ``spinor weight''.  Counting $\lambda_i$ as weight +1 and $\bar\lambda_i$ as $-1$, then the amplitude has 
weight $+4$ for 
a negative helicity graviton and $-4$ for a positive helicity graviton.

We define the theory starting with the the usual three-point amplitudes of Einstein 
gravity:\footnote{We remove a factor of $i(\kappa/2)^{n-2}$ from the $n$-point amplitude.}
\begin{eqnarray}
V_3( 1^-, 2^- , 3^+)&=& { \spa{1}.{2}^6 \over \spa1.3^2\spa3.2^2   }\;,
 \nonumber 
 \\
V_3( 1^+, 2^+, 3^-)&=& { \spb{1}.{2}^6 \over \spb1.3^2\spb3.2^2   }\;,
 \nonumber \\
V_3( 1^+, 2^+ , 3^+)&=&V_3( 1^-, 2^- , 3^-)=0\;.
\end{eqnarray}
These amplitudes have the correct spinor weight and are quadratic in the momenta.   
These amplitudes are only
defined for complex momenta. For an on-shell three-point amplitude the condition $k_1+k_2+k_3=0$ demands
$k_1\cdot k_2 =0$ etc.  For real momenta this implies $\spa{i}.{j}=\spb{i}.{j}=0$ and the vertices are all zero. However if we consider 
complex momenta then we can have $\lambda_1\sim \lambda_2 \sim \lambda_3$ but $\spb{i}.j \neq 0$.

The tree amplitudes for Einstein gravity can be computed recursively starting 
from these~\cite{Bedford:2005yy,Cachazo:2005ca,BjerrumBohr:2005jr}. We show that a similar construction can be used for an extended theory.

We extend this theory by adding additional three-point amplitudes which are of higher power in momenta.  To be non-trivial, 
these three-point amplitudes must either be functions of $\spa{i}.j$ or $\spb{i}.j$ exclusively.  The simplest polynomial amplitudes arise 
with six powers of momenta and are 
\begin{eqnarray}
V^{\alpha}_3( 1^-,  2^- , 3^-) &=&  {\alpha} { \spa{1}.{2}^2 \spa2.3^2\spa3.1^2   }\;,
 \nonumber  \\
V^{\alpha}_3( 1^+, 2^+, 3^+) &=& {\alpha} { \spb{1}.{2}^2  \spb2.3^2\spb3.1^2   }
\end{eqnarray}  
where ${\alpha}$ is an arbitrary constant.   We also have
\begin{equation}
V^{\alpha}_3( 1^-,  2^- , 3^+)=V^{\alpha}_3( 1^+,  2^+ , 3^-)=0\;,
\end{equation}
there being no polynomial function with the correct spinor and momentum weight.  {These are
essentially the unique choice for a three-point amplitude~\cite{Benincasa:2007xk}. }
\begin{figure}[ht]
   \includegraphics{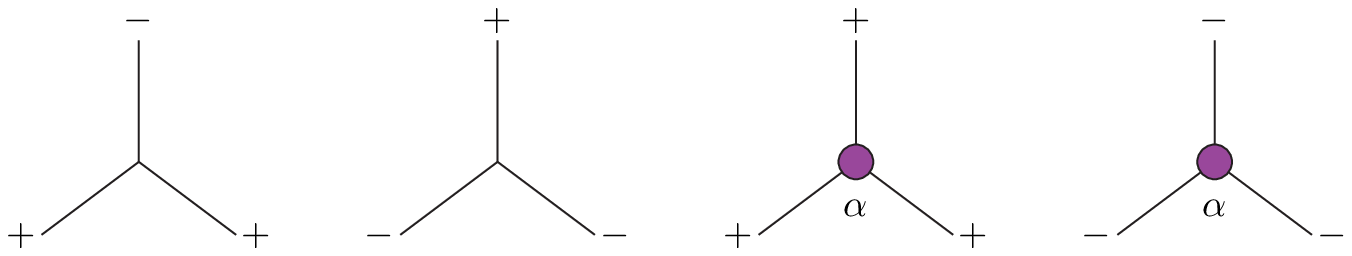}
    \caption{The non-zero three-point amplitudes}
    \label{fig:fivepointbubbles}
\end{figure}

The amplitudes in this theory can be expanded as a power series in $\alpha$,
\begin{equation}
M_n(1,\cdots , n) =  \sum_{r=0}  {\alpha}^r  M_n^{(r)}(1,\cdots , n)
\end{equation}
where $M_n^{(0)}$ is the Einstein gravity amplitude. 
Here we focus on the $r=1$ part of the extended theory.  This being the leading deformation of the theory from Einstein gravity. 

The theory we are considering would arise using field theory methods from the Lagrangian
\begin{equation}
L= \int d^D x \sqrt{-g} (  R +  C_{\alpha}  R_{abcd}R^{cdef}R_{ef}{}^{ab}   )
 \end{equation}
where $C_{\alpha}=\alpha/60$. 
However we note that to do so would involve determining increasingly complicated $n$-point vertices as the Lagrangian is expanded in the graviton field.      
As we will see the three-point amplitudes are sufficient to completely determine the $S$-matrix.

The key element is that  the entire $S$-matrix is determined from these vertices if we  demand that the amplitudes 
factorise on simple poles. 
Specifically, for any partition of the external legs into two sets, $\{k_{L_1},k_{L_2}\cdots,K_{L_l}\}$ and $\{k_{R_1},k_{R_2}\cdots,k_{R_m}\}$ 
with $l+m=n$ and $l,m\ge 2$,
if $K =\sum_{j=1}^l k_{L_j}$, then
when $K^2\longrightarrow 0$ the amplitude is singular with the simple pole being
\begin{equation}
M_{n}^{\tree}\ \mathop{\longrightarrow}^{K^2 \rightarrow 0}
\sum_{\lambda=\pm} \Biggl[ M_{l+1}^{\tree}  \big(k_{L_1}, \ldots,
k_{L_l}, -K^\lambda\big)  
\, {i \over K^2} \, M_{m+1}^{\tree} \big(K^{-\lambda}, k_{R_1}, \ldots,
    k_{R_m} \big) \Biggr] \;.
\end{equation}

We can excite the pole in $K^2$ by shifting to complex momenta and applying methods of complex analysis.
There are two shifts which we use to generate the $S$-matrix.  Firstly there is the original 
Britto-Cachazo-Feng-Witten (BCFW) shift~\cite{Britto:2005fq},
\begin{equation}
\lambda_i \longrightarrow \lambda_i+z\lambda_j \;,\;\;  
\bar\lambda_j \longrightarrow \bar\lambda_j- z\bar\lambda_i \;.
\label{eq:bcfwshift}
\end{equation}
For Einstein gravity this shift is sufficient to generate the tree level $S$-matrix~\cite{Benincasa:2007qj}.
Additionally we can use the Risager  shift~\cite{Risager:2005vk},
\begin{eqnarray}
\lambda_i \longrightarrow \lambda_i +z\spb{j}.{k} \lambda_{\eta} \;,
 \nonumber \\
\lambda_j \longrightarrow \lambda_j +z\spb{k}.{i} \lambda_{\eta}\;,
 \nonumber \\
\lambda_k \longrightarrow \lambda_k +z\spb{i}.{j} \lambda_{\eta}\;,
\label{eq:kaspershift}
\end{eqnarray}
where $\lambda_{\eta}$ is an arbitrary spinor.
Both shifts change the momenta to be functions of $z$ whilst leaving all momenta null and preserving overall momentum conservation.  We need
both shifts to construct the $S$-matrix for the extended theory.  By considering the integral
\begin{equation}
\int_{\gamma} { M(z) \over z }
\end{equation} 
where $\gamma$ is a closed contour, 
{\it provided} $M(z)$ vanishes at infinity the unshifted amplitude, $M(0)$,  can be obtained from the singularities in the amplitude. 
These occur at points $z_i$ where $K_i^2(z)=0$. At these points,
\begin{equation}
K_i^2(z)= -{ ( z-z_i) \over z_i} \times K_i^2(0)
\end{equation}
and we obtain,
\begin{equation}
M_n^{\tree} (0) \; = \; \sum_{i,\lambda} {M^{\tree,\lambda}_{l_i+1}(z_i)
  {i\over K_i^2(0)}M^{\tree,-\lambda}_{m_i+1}(z_i)},
\label{RecursionTree}
\end{equation}
where the summation over $i$ is only over factorisations where there are  shifted legs on both sides of the pole. This is the on-shell
recursive expression of~\cite{Britto:2005fq}. 
Note that if $M(z)$ does not vanish at infinity this does not imply factorisation is insufficient to determine the amplitude but
only that that particular shift can not be used to engineer the amplitude. 

Expressions obtained from~(\ref{RecursionTree})  are not manifestly symmetric as the choice of shift legs breaks crossing symmetry, however
symmetry is restored in the sum.  This is a highly non-trivial check that the amplitude has been computed successfully. 

\section{Generating the amplitudes}

In this section we give some of the details of the process of generating  the 
leading $\alpha$ contribution to the $S$-matrix. 

{\it Four-Point Amplitudes: }  
The three-point amplitudes are our inputs so the first outputs are the four-point amplitudes. There are three independent helicity configurations,
\begin{equation}
M_4( 1^+,2^+,3^+,4^+) \;, \;\;
M_4( 1^-,2^+,3^+,4^+)\; , \;\;
M_4( 1^-,2^-,3^+,4^+)\;  .
\end{equation}
Of these the first two are vanishing in Einstein gravity with only the last being non-zero: which is consequently termed the
``Maximally-Helicity-Violating''  (MHV) amplitude. 
For $M_4^{(1)}$ the reverse is true: $M_4^{(1)}( 1^-,2^-,3^+,4^+) =0$ since there are no possible factorisations,  while $M_4^{(1)}( 1^+,2^+,3^+,4^+)$
and $M_4^{(1)}( 1^-,2^+,3^+,4^+)$ are non-zero.

\begin{figure}[h]
   \includegraphics{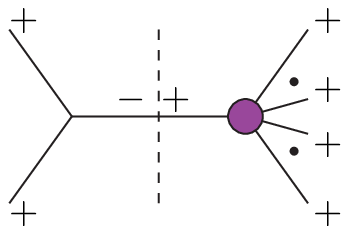}
    \caption{Factorisations of the $n$-point all-plus}
    \label{fig:allplusfactorisations}
\end{figure}

\begin{figure}[h]
\includegraphics{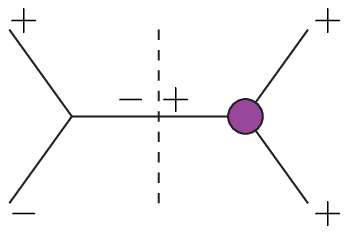}
    \caption{Factorisations of the four-point single minus amplitude}
    \label{fig:fourpointminus}
\end{figure}

The factorisations of the $n$-point all-plus amplitude  
are shown in fig.~\ref{fig:allplusfactorisations},
and the factorisations of the four-point single minus amplitude are shown on fig.~\ref{fig:fourpointminus}.

These factorisations can be excited using either of the shifts in~(\ref{eq:bcfwshift}) and~(\ref{eq:kaspershift}).
In the all-plus case only the second results in 
an amplitude with the correct symmetries. This in indication that~(\ref{eq:bcfwshift}) yields a shifted all-plus amplitude that
does not vanish at infinity. 
Conversely, for the single minus amplitude we must use the BCFW shift. 
Performing the shifts and evaluating the amplitudes we obtain
\begin{eqnarray}
{M}^{(1)}_4(1^+,2^+,3^+,4^+) &=& 10 
\left( { s t \over \spa1.2\spa2.3\spa3.4\spa4.1 } \right)^2 { stu }\,,
\nonumber
\\
{M}^{(1)}_4(1^-,2^+,3^+,4^+) &=& 
\left( { \spb2.4^2  \over \spb1.2\spa2.3\spa3.4\spb4.1 } \right)^2 
{ s^3 t^3 \over   u } \,.
\end{eqnarray}
The other non-zero amplitudes are available by conjugation. For the all-plus amplitude the recursion generates terms that contain the arbitrary spinor 
$\lambda_{\eta}$, however the sum of terms is independent of $\lambda_{\eta}$ and simplifies to the above.  
{These four-point amplitudes due to a $R^3$ term have been computed using field theory methods
long ago~\cite{vanNieuwenhuizen:1976vb}. These amplitudes vanish to all orders in a supersymmetric theory: a fact used show
supergravity was two-loop ultra-violet finite~\cite{Grisaru:1976nn,Tomboulis:1977wd}.
 The above expressions are in a spinor helicity basis  but agree once this is accounted for.  
In~\cite{Cohen:2010mi} these four-point amplitudes were also obtained using a ``all-line recursion'' technique where all legs have shifted momenta.  
}
These expressions also appear as the  UV infinite
 pieces of both  
two-loop gravity in four dimensions~\cite{Dunbar:2017qxb,Bern:2017puu} and one-loop gravity in six dimensions~\cite{Dunbar:2002gu}.

\noindent
{\it Five-Point Amplitudes:}
As before the shift ~(\ref{eq:kaspershift}) yields an all-plus amplitude that is independent of $\lambda_{\eta}$ and has full crossing symmetry:
\begin{equation}
{M}^{(1)}_5(1^+,2^+,3^+,4^+,5^+) =  \left( \sum_{P_6}  T^A_{(1,2,3),(4,5)}+\sum_{P_3} T^B_{(1,2,3),4,5} \right)
\label{eq:rcubedfivepoint}
\end{equation}
where
\begin{eqnarray}
T^A_{(1,2,3),(4,5)}= 10{ \spb1.4 \over \spa1.4 } 
{ \spb5.3  \spb5.2 \over \spa1.{\eta}^2\spa4.{\eta}  }{ \spb2.3^2
\over  \spa4.5 }\times 
  [5|K_{14}|\eta\ra 
 [2|K_{14}|\eta \ra 
 [3|K_{14}|\eta \ra\,,  
\end{eqnarray}
\begin{eqnarray}
T^B_{(1,2,3),4,5}=-10  { \spb1.4 \spb1.5 \spb2.3 [1|K_{23}|\eta\ra^2 [5|K_{23}|\eta\ra  [4|K_{23}|\eta\ra \over \spa2.3 \spa2.\eta^2\spa3.\eta^2  }
 {  \spb4.5 \over  \spa4.5   }
\end{eqnarray}
and $P_3$ denotes summation over the three cyclic permutations of legs 1,2 and 3. $P_6$ denotes the three permutations of $P_3$ together with 
interchange of legs 4 and 5.
The $\lambda_{\eta}$ independence of ${M}^{(1)}_5(1^+,2^+,3^+,4^+,5^+)$ is not manifest. 

\begin{figure}[h]
\includegraphics{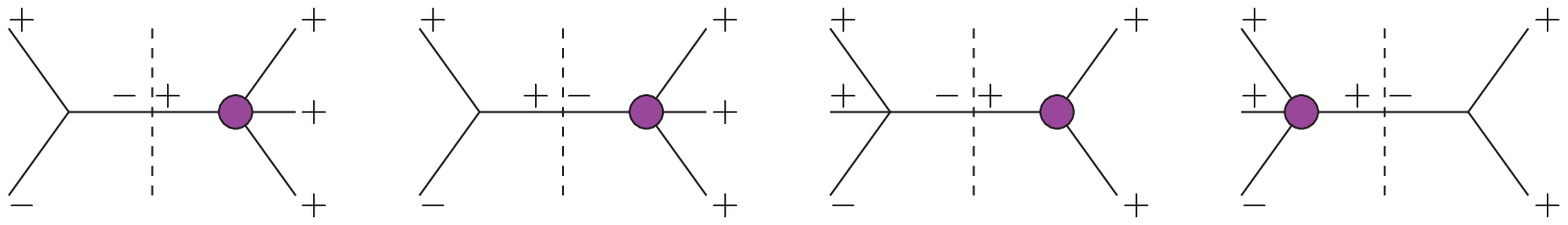}
    \caption{Factorisations of the five-point single minus amplitude}
    \label{fig:fivepointminus}
\end{figure}

The factorisations of the five-point single minus amplitudes are more varied as shown on fig.~\ref{fig:fivepointminus}.
Using the BCFW shift on $(\bar\lambda_1,\lambda_2)$ we obtain the amplitude 
\begin{eqnarray}
&M^{(1)}_5&(   1^-,   2^+,3^+,  4^+,5^+)=
{10\over \spb1.2^2 }
\left(  \prod_{i,j=2,3,4,5,i<j} \spb{i}.{j} \right) 
\left(
{ \spa1.5 \over \spb1.5}  { \spb2.5^3 \over  \spa3.4 }
+{ \spa1.3 \over \spb1.3}  { \spb2.3^3 \over  \spa4.5 }
+{ \spa1.4 \over \spb1.4}  { \spb2.4^3 \over  \spa5.3 }
\right)
\nonumber 
\\
&+&{\spa1.2^2  \over   \spa3.4\spa3.5\spa4.5   \prod_{i=3,4,5}  \spa1.{i} }  \left( 
{\spb2.3^5\spb4.5  \spa1.3^5 \over \spa{2}.3   } 
+{\spb2.4^5\spb5.3  \spa1.4^5 \over \spa{2}.4   } 
+{\spb2.5^5\spb3.5  \spa1.5^5 \over \spa{2}.5   } 
\right)
\nonumber
\\
&+& 
 {  1\over   \spa1.2^2 \spa3.4\spa3.5\spa4.5   } 
\Biggl( 
{ \spb2.3 \spb4.5^5 \spa1.5^3 \spa1.4^3  \over  \spa2.3 }  +
{ \spb2.4 \spb5.3^5 \spa1.3^3 \spa1.5^3  \over  \spa2.4 }  
\nonumber
\\
&+&{ \spb2.5 \spb3.4^5 \spa1.4^3 \spa1.3^3  \over  \spa2.5 }  
\Biggr)\;.
\end{eqnarray}

\begin{figure}[h]
\includegraphics{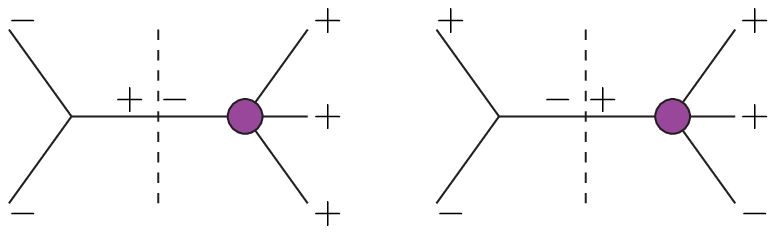}
    \caption{Factorisations of the five-point MHV amplitude}
    \label{fig:fivepointmhv}
\end{figure}

The five-point MHV amplitude is non-zero. 
The non-zero factorisations of the amplitude are shown in fig.~\ref{fig:fivepointmhv}.

This amplitude  can be obtained using a BCFW shift of either the two negative helicity legs or of a negative-positive pair. 
Shifting the two negative legs generates the expression (using only the second factorisation of fig.~\ref{fig:fivepointmhv}),  
\begin{eqnarray}
& &M_5^{(1)} (1^-,2^-,3^+,4^+,5^+)=
-s_{34} {\spa1.5\over  \spb1.5} { \spb3.4^2\spb3.5^3\spb4.5^3\over \spb1.2^2 \spb2.3\spb2.4 }
\nonumber 
\\
& & -s_{45} {\spa1.3\over  \spb1.3} { \spb4.5^2\spb4.3^3\spb5.3^3\over \spb1.2^2 \spb2.4\spb2.5 }
-s_{53} {\spa1.4\over \spb1.4} { \spb5.3^2\spb5.4^3\spb3.4^3\over \spb1.2^2 \spb2.5\spb2.3 }\;.
\end{eqnarray}

This completes the set of five-point amplitudes.  We can continue in this way generating the tree-level $S$-matrix.  
We have made available $M_n^{(1)}$ for $n \leq 7$ in {\tt Mathematica} format at
{\tt http://pyweb.swan.ac.uk/~dunbar/Smatrix.html}.
The amplitudes have been generated up to $n=8$ and have the correct symmetries, are $\eta$-independent and have the correct leading soft-limits.

We have evaluated amplitudes in a $R+\alpha R^3$ theory.  In ref.~\cite{Broedel:2012rc} amplitudes in Yang-Mills theory extended by $F^3$ terms were studied. Then using
double copy techniques and the KLT relations~\cite{Kawai:1985xq} graviton scattering amplitudes were derived upto $n=6$.  
As noted in~\cite{Broedel:2012rc} these correspond
to amplitudes in a $R+\alpha R^3+\sqrt{\alpha}R^2\phi$ theory. The four-point  amplitudes in the two theories are proportional~\cite{Cohen:2010mi,Broedel:2012rc} but 
beyond four-point the two sets of amplitudes are functionally different.  The all-plus amplitude in the two theories remain proportional for $n>4$  with
\begin{equation}
M^{(1),R^3+R^2\phi}_n(1^+,2^+,\cdots n^+) = \frac{5}{2} M^{(1),R^3}_n(1^+,2^+,\cdots n^+)
\label{eq:proportional}
\end{equation} 
and we confirm this for $n\leq 7$. 
\section{Soft Limits}

Graviton scattering amplitudes are singular as a leg becomes
soft. Weinberg~\cite{Weinberg} many years ago presented the leading
soft limit. 
If we parametrise the momentum of the $n$-th leg as $k_n^\mu = t \times k_s^\mu$ then  
in the
limit $t\longrightarrow 0$ the singularity in the $n$-point amplitude is
\begin{equation}
M_n \longrightarrow {1\over t} \times S^{(0)} \times M_{n-1} +O(t^0)
\label{softone}
\end{equation}
where $M_{n-1}$ is the {\it n--}1-point amplitude.    The soft-factor
$S^{(0)}$ is
universal and  Weinberg showed that (\ref{softone}) does not receive corrections in loop amplitudes.
 
\def\pol{\eps}
\def\sand#1.#2.#3{%
\left\langle\smash{#1}{\vphantom1}^{-}\right|{#2}%
\left|\smash{#3}{\vphantom1}^{-}\right\rangle}
\def\sandp#1.#2.#3{%
\left\langle\smash{#1}{\vphantom1}^{-}\right|{#2}
\left|\smash{#3}{\vphantom1}^{+}\right\rangle}
\def\sandpp#1.#2.#3{%
\left\langle\smash{#1}{\vphantom1}^{+}\right|{#2}%
\left|\smash{#3}{\vphantom1}^{+}\right\rangle}
\def\sandmm#1.#2.#3{%
\left\langle\smash{#1}{\vphantom1}^{-}\right|{#2}%
\left|\smash{#3}{\vphantom1}^{-}\right\rangle}

Recently it has also been proposed \cite{White:2011yy,Cachazo:2014fwa,He:2014laa} that the sub-leading and sub-sub-leading terms are also universal.  
This can be best exposed, when a positive helicity leg becomes soft,  by
setting
\begin{equation}
\lambda_n = t \times \lambda_s \;, \;\;\; 
\bar\lambda_n =  \bar\lambda_s \;.
\label{BasicSoft}
\end{equation}
In the $t\longrightarrow 0$ limit the amplitude has $t^{-3}$ singularities. At tree level the amplitudes 
satisfy soft-theorems~\cite{Cachazo:2014fwa} whereby their 
 behaviour as $t\longrightarrow 0$ is
\begin{equation}
\Mtree_n 
= S_t\Mtree_{n-1} +O(t^0) =\bigg(  {1\over t^3} S^{(0)} +{1\over t^2} S^{(1)} +{1\over t
} S^{(2)} \biggr) \Mtree_{n-1} +O(t^0)
\label{eq:softtheorem}
\end{equation}
where, for a positive helicity-leg becoming soft~\cite{Cachazo:2014fwa,Bern:2014oka,Broedel:2014fsa} 
\begin{eqnarray}
S^{(0)}&=&  -\sum_{i=1}^{n-1} { \spb{s}. i  \spa{i}.{\alpha} \spa{i}.\beta\over
  \spa{s}. i \spa{s}.{\alpha} \spa{s}.\beta }\;,
\\
S^{(1)} &=&  -\frac{1}{2} \sum_{i=1}^{n-1}    { \spb{s}.i \over \spa{s}.i }  
\left(   { \spa{i}.\alpha \over \spa{s}.\alpha }+{ \spa{i}.\beta
    \over \spa{s}.\beta } 
\right)
\bar\lambda_s^{\dot a} { \partial \over \partial \bar\lambda_i^{\dot a} }\;,
\\
S^{(2)} &=& \frac{1}{2} \sum_{i=1}^{n-1} { \spb{i}.s \over \spa{i}.s }
\bar\lambda_s^{\dot a} \bar\lambda_s^{\dot b}
{ \partial \over \partial \bar\lambda_i^{\dot a} } 
{ \partial \over \partial \bar\lambda_i^{\dot b} }\;.
\end{eqnarray}

The proof of the soft theorems follows
from Ward identities of extended Bondi, van der Burg, Metzner and
Sachs (BMS) symmetry~\cite{Bondi:1962px}.  Although exact for tree level amplitudes 
these receive loop corrections~\cite{Bern:2014oka,He:2014bga,Alston:2012xd}. 

{Whether the soft theorems extend beyond Einstein gravity has been examined before. In particular the leading 
soft behaviour can often be used as a check upon amplitudes such, e.g. in \cite{Broedel:2012rc}.  The leading and sub-leading
limits were shown to hold for a $R^3$ insertion in~\cite{Bianchi:2014gla}. 
Here we examine the amplitudes and, in particular, test the sub-sub-leading soft behaviour.}

We can summarise the behaviour of the leading amplitudes, $M^{(1)}_n$, simply  by stating:

\begin{center}
{\bf All the amplitudes calculated satisfy the soft limits of 
\break 
(\ref{eq:softtheorem}) 
 up to and including the sub-sub-leading term.}  
\end{center}

We have verified this for all helicity amplitudes up to $n=8$. 
Note: to check~(\ref{eq:softtheorem}) one must implement momentum conservation consistently between the $n$-point amplitudes and the $n-1$-point amplitudes
which in essence specifies how the point $t=0$ is approached.  These are several ways to do this. We have followed the prescription
of~\cite{Cachazo:2014fwa} but alternative
implementations are possible~\cite{Bern:2014oka,Broedel:2014fsa}. 

\noindent 
In principle we could have found a behaviour of the form
\begin{equation}
M_n^{(1)} \longrightarrow   S_t  M_{n-1}^{(1)} +S_t^{\alpha} M_{n-1}^{(0)}  +R_n
\end{equation}
where $S_t^{\alpha}$ would be an $\alpha$ correction to the soft functions and $R_n$ is a non-factorising term. 
In terms of this we find $S_t^{\alpha}=R_n=0$. 
Since the theory we are considering is higher derivative it is not surprising that the leading and sub-leading parts of $S_t^{\alpha}$ 
vanish however it is interesting that the vanishing continues for the sub-sub-leading - unlike the loop corrections to Einstein gravity.

Incidentally as a consequence of eq.(\ref{eq:proportional}) the amplitude $M^{(1),R^3+R^2\phi}_n(1^+,2^+,\cdots n^+)$ also satisfies the soft theorems to sub-sub leading level.

\section{Other Theories}

We have chosen to extend gravity using a three-point vertex and use a diagrammar approach whereby we only consider the on-shell amplitudes.  
There is, of course, complementarity between this approach and that of Lagrangian based field theory.  The single choice of three-point amplitude corresponds
to the single $R^3$ field density that  affects on-shell amplitudes.  This makes the extended $S$-matrix simply depend upon the single parameter $\alpha$.

If we were to deform Einstein gravity by an additional four-point amplitude then there are more choices consistent with symmetry and 
spinor weight, e.g. we could have
\begin{eqnarray}
M_4 (1^+,2^+,3^+,4^+) =
&\alpha_1 & ( \spa1.2^4\spa3.4^4+\spa1.3^4\spa2.4^4+\spa1.4^4\spa2.3^4 )
\nonumber
\\
+&\alpha_2 & (  \spa1.2\spa2.3\spa3.4\spa4.1 +\rm{permutations} )^2
+\cdots
\end{eqnarray}
From a field theory perspective this freedom corresponds to the observation that there are multiple $R^4$
tensors that contribute to on-shell amplitudes~\cite{Fulling:1992vm}.

\begin{figure}[h]
\includegraphics{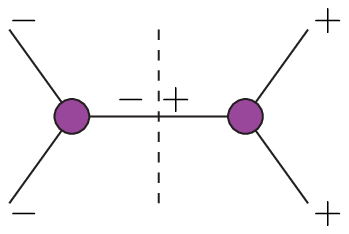}
    \caption{Factorisations of the four-point MHV amplitude at $\alpha^2$.}
    \label{fig:alphasquared}
\end{figure}

The same issue arises when we consider the further expansion in $\alpha$.  If we consider 
$M_4^{(2)}(1^-,2^-,3^+,4^+)$ there is a single factorisation as shown in fig.~\ref{fig:alphasquared}.  The amplitude
\begin{equation}
M^{(2)}_4 (1^-,2^-,3^+,4^+) =  \spa1.2^4\spb3.4^4 \left(  {tu+\beta s^2 \over s } \right)
\end{equation}
has the correct factorisation for any choice of $\beta$.  This ambiguity means we also have to specify the four-point amplitude to 
determine the $S$-matrix. 
In the diagrammar approach this ambiguity
arises due to the existence of a polynomial function with the correct symmetries and spinor and momentum weight.  
From a field theory perspective, additional counterterms can contribute to this amplitude. Specifically,  we could deform the theory via
\begin{equation}
R \longrightarrow R + C_\alpha  R^3  + C_\beta D^2 R^4
\end{equation}
and the four-point amplitude is only specified once $C_{\alpha}$ and $C_{\beta}$ are determined. 

\section{Conclusion}

We have constructed the (leading part) of the $S$-matrix of an extended theory of gravity starting from three-point amplitudes and only demanding
factorisation. The theory is extended by the  addition of amplitudes which  are polynomial in momentum, thus implicitly imposing locality and 
unitarity on the $S$-matrix. We also require the amplitudes to have the correct spinor helicity as appropriate for massless particles.  The $S$-matrix
is then generated entirely from on-shell amplitudes by demanding factorisation. Specifically, we have extended the theory by the addition 
of three-point amplitudes which, from a field theory perspective, corresponds to introducing $R^3$ terms.  This $S$-matrix differs from that obtained by 
applying double copy or KLT techniques to a   $F^3$ extension of Yang-Mills.

Beyond the leading part, polynomial amplitudes exist at higher point and these must be specified to fully determine the $S$-matrix. 
Consistency of this approach and a field theoretic approach beyond leading order requires a correspondence between these polynomial amplitudes and 
the counter terms contributing to on-shell amplitudes.

We find that these amplitudes satisfy the same soft theorems as the tree amplitudes of Einstein gravity up to and including the sub-sub leading terms. It is interesting that these theorems are 
robust to deformations of Einstein gravity even at the sub-sub-leading level particularly given the link to BMS symmetry
which plays an important role in the recent understanding of black hole soft hair~\cite{Hawking:2016msc}.

\section{Acknowledgements}

This work was supported by STFC grant ST/L000369/1.  GRJ was supported by STFC grant ST/M503848/1.
JHG was supported  by the College of Science (CoS)
Doctoral Training Centre (DTC) at Swansea University.

 \appendix

\end{document}